\documentclass[conference]{IEEEtran}
\IEEEoverridecommandlockouts
\usepackage{cite}
\usepackage{amsmath,amssymb,amsfonts}
\usepackage[numbers]{natbib}
\usepackage[colorlinks,citecolor=black,linkcolor=black]{hyperref}
\usepackage{subfig}  
\usepackage{algorithm}
\usepackage{algorithmic}
\usepackage{graphicx}
\usepackage{textcomp}
\usepackage{xcolor}
\usepackage{makecell}
\def\BibTeX{{\rm B\kern-.05em{\sc i\kern-.025em b}\kern-.08em
    T\kern-.1667em\lower.7ex\hbox{E}\kern-.125emX}}
\begin{document}

\title{DifFaiRec: Generative Fair Recommender with Conditional Diffusion Model
}

\author{
\IEEEauthorblockN{Zhenhao Jiang}
\IEEEauthorblockA{\textit{The Chinese University of Hong Kong, Shenzhen}\\
Guangdong, China \\
222041010@link.cuhk.edu.cn}
\and
\IEEEauthorblockN{Jicong Fan*}
\IEEEauthorblockA{\textit{The Chinese University of Hong Kong, Shenzhen}\\
Guangdong, China \\
fanjicong@cuhk.edu.cn}
\thanks{*Corresponding author}
}

\maketitle

\begin{abstract}
Although recommenders can ship items to users automatically based on the users' preferences, they often cause unfairness to groups or individuals. For instance, when users can be divided into two groups according to a sensitive social attribute and there is a significant difference in terms of activity between the two groups, the learned recommendation algorithm will result in a recommendation gap between the two groups, which causes group unfairness. In this work, we propose a novel recommendation algorithm named \underline{Dif}fusion-based \underline{Fai}r \underline{Rec}ommender (DifFaiRec) to provide fair recommendations. DifFaiRec is built upon the conditional diffusion model and hence has a strong ability to learn the distribution of user preferences from their ratings on items and is able to generate diverse recommendations effectively. To guarantee fairness, we design a counterfactual module to reduce the model sensitivity to protected attributes and provide mathematical explanations. The experiments on benchmark datasets demonstrate the superiority of DifFaiRec over competitive baselines. 
\end{abstract}

\begin{IEEEkeywords}
Recommender System, Group Fairness, Diffusion Model, Counterfactual Module
\end{IEEEkeywords}

\section{Introduction}
Recommendation algorithms can improve the efficiency of interactions between users and items, have wide applications in e-commerce and social media platforms \cite{ma2018entire,NEURIPS2019_0fc170ec,zhou2019deep,zhu2022personalized,fan2024neuronenhanced}, and have been changing our life and habits explicitly or implicitly. However, recommenders can lead to unfairness about sensitive social attributes such as gender and age, which requires us to face with caution \cite{ge2021towards,beutel2019fairness}. Recently, research on the fairness of recommendation algorithms has attracted the attention of many scholars \cite{beutel2019fairness,fu2020fairness,polyzou2021faireo} and a few fair recommendation algorithms have been proposed. 

Unlike traditional Click-Through Rate (CTR) prediction recommendation algorithms, generative recommendation, often based on deep generative models, directly gives the estimated list of the entire set of candidate products using the learned interaction distribution \cite{li2022fairgan,wang2023generative,yuan2019simple,luo2020deep}. This list-wise generation is more likely to give a recommendation list that the user prefers and perceive the interaction information between users and items and that among items in the candidate set, which makes generative recommender achieve a remarkable success \cite{deldjoo2021survey,li2022fairgan}. Classical deep generative models include variational autoencoder \cite{kingma2019introduction}, flow-based generative model \cite{ho2019flow++}, and generative adversarial networks \cite{goodfellow2020generative}, etc. Compared to these classical models, recently, diffusion models \cite{sohl2015deep} have shown astonishing generative ability and achieved state-of-the-art results in image generation \cite{tashiro2021csdi,croitoru2023diffusion}. This motivated several diffusion model based recommendation algorithms \cite{du2023sequential,li2023diffurec,wang2023diffusion}.

In addition, the existing fair recommenders have at least two optimization objectives \cite{biswas2021toward,qi2022profairrec,ge2022explainable}, namely fairness and accuracy. However, it is difficult to find Pareto optimality \cite{zhang2022pareto} for multi-objective optimization problems, and the trade-off between accuracy and fairness is hard to balance. This greatly limits the application of fair recommenders. Fortunately, we find that diffusion models provide a possible solution to this issue. We can employ Bayesian ideology to compress objectives of fairness and accuracy into one. In this way, we can only focus on one optimization goal, reducing the difficulty of optimization.

In this work, we propose a diffusion-based fair recommender (DifFaiRec) for group-fair recommendations.
DifFaiRec contains two modules. The first one is a counterfactual module that maps each user to its different group to obtain a counterfactual user, which is to ensure group fairness. The second module is a conditional diffusion model that is conditioned on the counterfactual module, reconstructs the observed interactions, and predicts the unknown interactions in a generative manner.  
Our contributions are as follows.
\begin{itemize}
    \item We propose a novel fair recommendation algorithm DifFaiRec. It is the first diffusion model based fair recommender.
    \item We design a novel fairness strategy.
    It is built on counterfactual samples and can force the recommender to give fair recommendations. 
    \item We compress the two objectives (accuracy and fairness) into one and provide a mathematical analysis of why this method works. 
\end{itemize}
Experiments on two real datasets show that our DifFaiRec is both accurate and fair and outperforms various baselines. 

\section{Related Works}
\subsection{Group Fairness in Recommendation}
The fairness of recommendation systems can be divided into many domains according to different views such as individual vs group, user vs item, and so on \cite{wang2023survey}. Here, we focus on group fairness, which holds that recommendation performance should be fair among different groups. For fairness in recommendation, the groups of users are often defined according to basic social attributes like gender, age, and career. \cite{ekstrand2018exploring} first evaluated the response of collaborative filtering algorithms to attributes of social concern, namely gender on publicly available book ratings data. \cite{fu2020fairness} formalized the fairness problem as a 0–1 integer programming problem and invoked a heuristic solver to compute feasible solutions. \cite{islam2021debiasing} proposed a neural fair collaborative filtering method that considers gender bias in recommending career-related items using a pre-training and fine-tuning framework. \cite{li2021leave} designed two auto-encoders for users and items working as adversaries to the process of minimizing the rating prediction error. They enforced that the specific unique properties of all users and items are sufficiently well incorporated and preserved in the learned representations to provide fair recommendations. 

There are a few studies that explore the issue of group fairness from a very different perspective. \cite{li2021user} found that the level of activity of users could also cause unfairness. They provided a re-ranking approach under fairness metric based constraints to address this unfairness problem. \cite{wan2020addressing} studied a special issue on underrepresented market segments. For example, a male user who would potentially like one product may be less likely to interact with them if it is using a 'female' image. \cite{beutel2019fairness} designed a novel metric based on pairwise comparisons to maintain fairness in the rankings given by the recommender. 
\cite{polyzou2021faireo} formulated their method as a multi-objective optimization problem and considered the trade-offs between equal opportunity and recommendation quality. 

Different from the existing research on group fairness in recommendation, our method relaxes the fairness and accuracy objectives into one by using the characteristics of the diffusion model and the Bayesian ideology, that is, the fairness recommendation task is transformed from the existing multi-objective optimization problem to the single-objective optimization problem, which reduces the difficulty of solving.

\subsection{Generative Recommender} 
Generative recommenders are usually based on deep generative models such as variational auto-encoder (VAE) \cite{zhao2022normalized}, generative adversarial network (GAN) \cite{goodfellow2020generative}, and diffusion models \cite{sohl2015deep,ho2020denoising}. 

VAE consists of two parts: an encoder and a decoder. The encoder outputs the parameters of the latent distribution, such as mean and variance. The decoder takes a sample from the latent distribution and gives a reconstruction of input data \cite{dai2019diagnosing}. \cite{zhang2023vae} introduced a cross-domain recommendation based on VAE. It extracts information from within-domain and cross-domain and stresses user preference in specific domains. \cite{xie2021adversarial} designed a VAE model with adversarial and contrastive learning for sequential recommendations. The model can learn more personalized and significant attributes. 
    
GAN consists of a generator and a discriminator, and is trained in an adversarial way \cite{karras2020analyzing}. \cite{li2022fairgan} presented a fairness-aware GAN for a recommendation system with implicit feedback. \cite{wu2019pd} proposed a novel GAN-based recommender to give a set of diverse and relevant items with a designed kernel matrix that considers both co-occurrence of diverse items and personal preference.

Similar to VAE and GAN, diffusion models \cite{sohl2015deep,ho2020denoising}, to be detailed in the next section, are also applicable and useful in recommendation systems \cite{du2023sequential,li2023diffurec,wang2023diffusion}.
For instance, \cite{wang2023diffusion} proposed a diffusion recommender model that learns the generative process in a denoising manner and has improved performance on several benchmark datasets. It is worth noting that \cite{du2023sequential,li2023diffurec,wang2023diffusion} do not address the fairness problem in recommendation systems.

Large language model (LLM)-based recommendation systems have become a hot topic recently. \cite{liu2023chatgpt} proposed a ChatGPt-based recommendation system that uses ChatGPT to make recommendations based on a user's historical behavior by designing a special prompt. This paper explores the performance of large language models on recommendation systems.  \cite{chen2023large} proposed the opportunities and challenges that the LLM-based recommendation system can encounter, and provides a broad idea for researchers. \cite{he2023large} proposed a zero-shot LLM-based conversation recommender, which can improve user retention and extract user interest more accurately.

Different from the existing generative recommenders, our method is based on the conditional diffusion model combining the counterfactual module to achieve a fair and accurate recommendation system. In addition, we give a mathematical explanation for the proposed method, which increases the interpretability of the model.

\section{Preliminary}\label{sec_pre}
Diffusion models (DMs) \cite{sohl2015deep} have recently attracted great attention in the fields of image and speech generation. Importantly, prompt-guided generation with conditional diffusion has fully demonstrated the controllability and flexibility of DMs. In this section, we briefly introduce the forward process, reverse process, training, and sampling of DMs \cite{ho2020denoising,kong2020diffwave}. 

\begin{itemize}
    \item \textbf{Forward process}~ The forward process takes an input sample $\mathbf{x}_0$ that follows the distribution $q(\mathbf{x}_0)$ and creates the latent samples {$\mathbf{x}_1,\ldots,\mathbf{x}_T$} by adding Gaussian noises step by step for $T$ times forming a Markov process \cite{ho2020denoising}:
    \begin{equation}
        q(\mathbf{x}_t|\mathbf{x}_{t-1}):=\mathcal{N}(\mathbf{x}_t;\sqrt{1-\beta_t}\mathbf{x}_{t-1},\beta_t\mathbf{I}),
    \end{equation}
    where $t\in\{1,...,T\}$ is the diffusion step, $\beta_t$ is the variance, and $\mathbf{I}$ is an identity matrix.
    \item \textbf{Reverse process}~ The reverse process is a denoising process starting at $p(\mathbf{x}_T)=\mathcal{N}(\mathbf{x}_T;\mathbf{0},\mathbf{I})$ \cite{hoogeboom2021argmax}. It aims to recover $\mathbf{x_0}$ from $\mathbf{x}_T$ according to
    \begin{equation}
        p_{\theta}(\mathbf{x}_{t-1}|\mathbf{x}_t):=\mathcal{N}(\mathbf{x}_{t-1};\mu_\theta(\mathbf{x}_t,t),\Sigma_\theta(\mathbf{x}_t,t)),
    \end{equation}
    where $\mu_\theta(\mathbf{x}_t,t)$ and $\Sigma_\theta(\mathbf{x}_t,t)$ are mean and covariance of Gaussian inferred by a network parameterized by $\theta$.
    \item \textbf{Training}~ The objective of DM is to optimize the variational bound on negative log-likelihood \cite{ho2020denoising}:
    \begin{equation}
    \begin{split}
        E[-\log(p_\theta(\mathbf{x_0}))]\leq E_q\left[-\log\left(\frac{p_\theta(\mathbf{x}_{0:T})}{q(\mathbf{x}_{1:T}|\mathbf{x}_0})\right)\right]\\
        =E_q\left[-\log(p(\mathbf{x}_T))-\sum_{t=1}^T\log\left(\frac{p_\theta(\mathbf{x}_{t-1}|\mathbf{x}_{t})}{q(\mathbf{x}_t|\mathbf{x}_{t-1})}\right)\right]
    \end{split}
    \end{equation}
    With some deduction and simplification, one can obtain the following  objective \cite{ho2020denoising}: 
    \begin{equation}
        E_{t,\mathbf{x}_t,\epsilon}\left[\Vert \epsilon-\epsilon_\theta(\mathbf{x}_t,t) \Vert^2_2\right],
    \end{equation}
    where $\epsilon\sim N(\mathbf{0},\mathbf{I})$ and $\epsilon_\theta$ is the predicted noise to determine $x_t$ from $x_0$.
    \item \textbf{Sampling} With the well trained $\theta$, DM samples $\mathbf{x}_0$ iteratively from $\mathbf{x}_T$ according to 
    \begin{equation}
        \mathbf{x}_{t-1}=\frac{1}{\sqrt{\alpha_t}}\left(\mathbf{x}_t-\frac{1-\alpha_t}{\sqrt{1-\bar{\alpha}_t}} \boldsymbol{\epsilon}_\theta\left(\mathbf{x}_t, t\right)\right)+\sigma_t \mathbf{z},
    \label{sampling}
    \end{equation}
    where $\alpha_t=1-\beta_t$ and $\bar{\alpha}_t=\prod_{s=1}^t \alpha_s$. More details can be found in \cite{ho2020denoising}.
\end{itemize}

\section{Fair Diffusion Recommender}
To take advantage of the generating ability and flexibility of the conditional diffusion model to address group unfairness, we design DifFaiRec that contains three components: 1) group vector builder; 2) counterfactual module; 3) diffusion model.

\subsection{Problem and Model Formulation}
Here, we state the problem of optimal ranking under group fairness constraint. We use $I^m$ and $U^n$ to denote the sets of $m$ items and $n$ users respectively. 
For $I^m$ and $U^n$, there is an interaction matrix denoted by $\mathbf{R}=[\mathbf{r}_1,\mathbf{r}_2,\ldots,\mathbf{r}_n]\in\mathbb{R}^{m\times n}$, where $\mathbf{r}_j$ denotes the $j$-th column of $\mathbf{R}$ and $r_{ij}$ is the rating given by user $j$ on item $i$. $r_{ij}=0$ means there is no interaction between user $j$ and item $i$. For convenience, we use a mask matrix $\mathbf{M}\in\{0,1\}^{m\times n}$ to denote whether the ratings in $\mathbf{R}$ are missing or not. We would like to construct an algorithm $\mathcal{A}$ that learns a prediction model $f_{\mathcal{A}}$ from $\mathbf{R}$ that is able to predict the unknown ratings accurately, namely, the following test error is sufficiently small:
\begin{equation}\label{prob_0}
E_{\text{test}}:=\frac{1}{mn-\sum_{ij}{M_{ij}}}\sum_{(i,j):M_{ij}=0}\left[\ell\left([f_{\mathcal{A}}(\mathbf{R})]_{ij},r_{ij}^\ast\right)\right],
\end{equation}
where $r_{ij}^\ast$ denotes the ground-truth rating, $\ell$ denotes a loss function such as the square loss, and $f_{\mathcal{A}}$ is conducted on $\mathbf{R}$ column-wisely, i.e., $f_{\mathcal{A}}(\mathbf{R})=\left[f_{\mathcal{A}}(\mathbf{r}_1),\ldots,f_{\mathcal{A}}(\mathbf{r}_n)\right]$. Note that $E_{\text{test}}$ is related to the training error and the complexity of $f_{\mathcal{A}}$.

Suppose the users are divided into two groups, denoted as $A$ and $B$, according to a sensitive attribute $s\in\{0,1\}$ such as gender.
We hope the prediction $\mathbf{v}$ given by $f_{\mathcal{A}}$ is fair for $A$ and $B$, i.e., the following prediction bias is sufficiently small:
\begin{equation}\label{eq_delta}
    \delta:=\left\vert P\left(f_{\mathcal{A}}(\mathbf{r})=\mathbf{v}\mid s=0\right)-P\left(f_{\mathcal{A}}(\mathbf{r})=\mathbf{v}\mid s=1\right)\right\vert.
\end{equation}
Simultaneously obtaining small $E_{\text{test}}$ and ${\delta}$ is difficult and there usually exists a trade-off between them. Moreover, achieving small ${\delta}$ is a challenging task because one has to estimate the probability in a high-dimensional space.

We achieve small $E_{\text{test}}$ and $\delta$ using a diffusion model conditioned on group vectors in a counterfactual manner. Let
\begin{equation}
  g(\mathbf{z})=f_{\mathcal{A}}(\mathbf{r}),  
\end{equation}
where $\mathbf{z}$ is some intermediate variable related to $\mathbf{r}$. $\mathbf{z}$ can be regarded as a feature representation vector of a user. Thus, the predicted ratings are given by $\mathbf{v}=g(\mathbf{z})$. If $\mathbf{z}$ is independent from $s$, then $\mathbf{v}$ is independent from $s$, which means $\delta=0$. To approximately achieve such a $\mathbf{z}$, we use a counterfactual method. The idea is as follows:
\begin{itemize}
    \item We represent groups $A$ and $B$ as two vectors $\mathbf{a}$ and $\mathbf{b}$ respectively.
    \item Given a user $j\in A$, we generate $\mathbf{z}_j'=\mathcal{C}(\mathbf{z}_j,\mathbf{b})$, where $\mathcal{C}$ is a counterfactual module. This $\mathbf{z}_j'$ corresponds to a counterfactual user in group $B$. Similarly, if $j\in B$, we let $\mathbf{z}_j'=\mathcal{C}(\mathbf{z}_j,\mathbf{a})$.
    \item The above construction means for a counterfactual user, $s=0$ and $s=1$ hold simultaneously. Thus, the value of $s$ has no (ideally) impact on $\mathbf{z}'$, which means $\mathbf{z}'$ is independent (ideally) from $s$.
    \item The prediction is then based on $g(\mathbf{z}')$ rather than $g(\mathbf{z})$:
    \begin{equation}
        \mathbf{v}=f_{\mathcal{A}}(\mathbf{r})=g(\mathbf{z}')=g(\mathcal{C}(\mathbf{z},\mathbf{y})),
        \label{eq9}
    \end{equation}
    where $\mathbf{y}\in\{\mathbf{a},\mathbf{b}\}$.
    Therefore, the prediction is fair to the two groups $A$ and $B$.
\end{itemize}
Fairness requires the model to give a similar prediction for the same user under different conditions. We can turn the requirement into an equivalent. If condition A becomes condition B for the same user, but nothing else is changed, the user becomes a counterfactual user. At this point, if the model's predicted rating for the counterfactual user remains the same, the requirement is satisfied. Further, if we directly input a counterfactual user to fit the interest of the real user, the model will treat both the counterfactual user and the real user alike. Because we made the model think that these two users have the same interests. In this way, we guarantee the requirement of fairness. \eqref{eq9} describes this idea.

In the following two sections, we introduce the design of group vectors $\mathbf{a}$ and $\mathbf{b}$ and the counterfactual module $\mathcal{C}$.



\subsection{Constructing Group Vectors}
To achieve group fairness, the first step is to determine the feature space of two groups that is good for us to distinguish the difference between them. 

One may classify users into two groups according to a sensitive attribute such as gender, where the label is a scalar chosen from $\{-1,1\}$. However, a scalar contains limited information about the group. Existing strategies often use embedding methods to represent a group as a low-dimensional dense vector \cite{tang2018personalized}. In fact, before conducting recommendations, the sensitive feature has already caused a gap between the ratings of the two groups. In other words, the training data already contains unfairness. Hence, we try two methods, mean pooling and principal component analysis (PCA) \cite{wold1987principal,abdi2010principal}, to build feature space based on the given ratings. They are able to quantify the gap or unfairness between two groups.

Mean pooling and PCA are operated along the user axis to obtain group vectors $\mathbf{a}$ and $\mathbf{b}$. Mean pooling is formulated as:
\begin{equation}
    \mathbf{a}=\frac{\Sigma_{j\in A}\mathbf{r}_j}{|A|},\quad \mathbf{b}=\frac{\Sigma_{j\in B}\mathbf{r}_j}{|B|}.
\end{equation}

For PCA, we partition the columns of $\mathbf{R}$ into two matrices $\mathbf{R}_A$ and $\mathbf{R}_B$ according to the sensitive attributes $s$, and let $\mathbf{C}_A$ and $\mathbf{C}_B$ be the covariance matrices computed from $\mathbf{R}_A$ and $\mathbf{R}_B$ respectively. We perform eigenvalue decompositions on $\mathbf{C}_A$ and $\mathbf{C}_B$, i.e.,
\begin{equation*}
    \mathbf{U}_A\boldsymbol{\Lambda}_A \mathbf{U}_A^\top=\mathbf{C}_A,\quad  \mathbf{U}_B\boldsymbol{\Lambda}_B\mathbf{U}_B^\top=\mathbf{C}_B,
\end{equation*}
where $\mathbf{U}_A=(\mathbf{u}_A^{(1)},\ldots,\mathbf{u}_A^{(m)})$, 
$\mathbf{U}_B=(\mathbf{u}_B^{(1)},\ldots,\mathbf{u}_A^{(m)})$, $\boldsymbol{\Lambda}_A=\text{diag}$
$(\lambda_A^{(1)},\ldots,\lambda_A^{(m)})$, $\boldsymbol{\Lambda}_B=\text{diag}(\lambda_B^{(1)},\ldots,\lambda_B^{(m)})$,
$\lambda_A^{(1)}\geq\cdots\geq\lambda_A^{(m)}$, and $\lambda_B^{(1)}\geq\cdots\geq\lambda_B^{(m)}$. Now we let the group vector be the first principal component, i.e.,
\begin{equation}
    \mathbf{a}=\mathbf{u}_A^{(1)},\quad \mathbf{b}=\mathbf{u}_B^{(1)}.
\end{equation}
Note that now $\mathbf{a}$ and $\mathbf{b}$ are the first basis vectors of column spaces of $\mathbf{R}_A^\top$ and $\mathbf{R}_B^\top$ respectively. They can be regarded as the feature vectors of the two groups.

\begin{figure*}[t]
	\centering
        \includegraphics[width=1\linewidth]{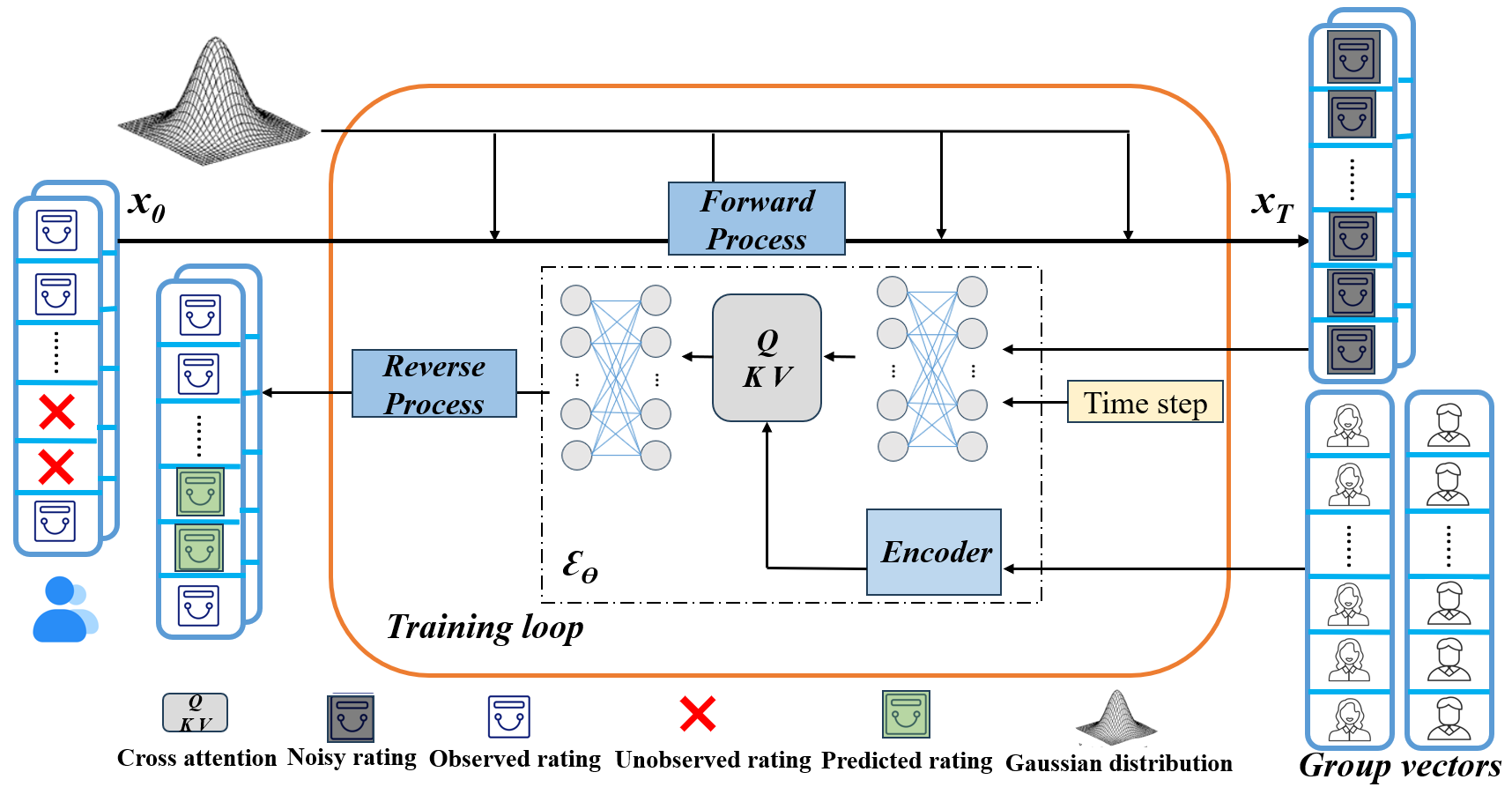}
        \caption{The flowchart of DifFaiRec. The original rating vector is $\mathbf{x}_0$. In the forward process, the vector is added with Gaussian noise $T$ times, becoming $\mathbf{x}_T$. In the reverse process, $\mathbf{x}_T$, the group vectors, and the time step are fed into DifFaiRec to estimate the noise, and then the missing ratings in $\mathbf{x}_0$ can be recovered after $T$ times of denoising.}
 \label{fig1}
\end{figure*}

\subsection{Counterfactual Module}
Here, we introduce a counterfactual module to keep fairness recommendations. Intuitively, if group fairness is satisfied, then the recommendation system has the same recommendation performance for the two groups. Further, if the sensitive characteristic $s$ of each individual in a group changes while the recommendation performance remains unchanged, it indicates that the algorithm is fair that is \cite{li2021user}:
\begin{equation}
f_{\mathcal{A}}(\mathbf{r};s=0)=f_{\mathcal{A}}(\mathbf{r};s=1),
\end{equation}
which is consistent with \eqref{eq_delta}.
However, given a user, the sensitive characteristic is fixed, and there is no real user who changes his/her sensitive characteristic. Therefore, we need to create a user who changes it, which is a counterfactual operation.

A simple method is to find a user who is similar to the user in another group as the research object but it is very difficult to find a one-to-one mapping for each user. Thus, we design a counterfactual module to map users to another group directly. Firstly, group vectors are transformed by a condition encoder:
\begin{equation}
    \Bar{\mathbf{g}} = \mathrm{Enc}(\mathbf{g}), \quad \mathbf{g}\in\{\mathbf{a},\mathbf{b}\},
\end{equation}
where Enc can be a multi-layer perception (MLP).

Then, an attention mechanism is implemented to run the counterfactual operation. Attention can be formulated as follows \cite{vaswani2017attention}:
\begin{equation}
\text{Atten}(query,key,value)=\text{Softmax}\left(\frac{\mathbf{\hat{q}}\cdot \mathbf{\hat{k}}^\top}{\sqrt{d}}\right)*\mathbf{\hat{v}},
\end{equation}
where $\mathbf{\hat{q}}=\mathbf{W}_q\cdot query,\mathbf{\hat{k}}=\mathbf{W}_k\cdot key, \mathbf{\hat{v}}=\mathbf{W}_v\cdot value$, $\mathbf{W}_q,\mathbf{W}_k,\mathbf{W}_v$ are parameter matrices and $d$ is the dimension of $\mathbf{\hat{q}}\cdot \mathbf{\hat{k}}^\top$. Here, the group vector is the query and the user vector is used as both key and value. 
Counterfactual module maps users in group $A$ to group $B$ with the help of attention: 
\begin{equation}
    \mathbf{z}'_j=\text{Atten}(\Bar{\mathbf{g}},\mathbf{z}_j,\mathbf{z}_j)\triangleq\mathcal{C}(\mathbf{z}_j,\mathbf{y}_j),
\end{equation}
where
\begin{equation}
\mathbf{y}_j=
    \begin{cases}
        \mathbf{a}\quad\text{if}~j\in B,\\
        \mathbf{b}\quad \text{if}~j\in A.
    \end{cases}
\end{equation}
The attention mechanism is able to adaptively mine the underlying connection between two vectors and output the correlation between them (attention score). But this is equivalent to imposing an equal correlation on each dimension of the vector, which is actually unreasonable. This is the reason for adding a condition encoder $D$ in the counterfactual module. The condition encoder can adaptively perform feature crossover, and then use attention to mine the counterfactual mapping relationship among users' ratings for different items. 

\subsection{Model Training}
In our algorithm $\mathcal{A}$, the model can predict the noise-matching term, and then denoise the samples via the following closed-form expression \cite{luo2022understanding}:
\begin{equation}
q(\mathbf{x_{t-1}}|\mathbf{x}_t,\mathbf{x_0}) \propto N(\mathbf{x_{t-1}};\Hat{\mu}(\mathbf{x}_t,\mathbf{x_0},t,\mathbf{y}),\sigma^2(t)\mathbf{I}),
\end{equation}
where $\mathbf{x}_0$ is sampled from $\mathbf{R}$, $\Hat{\mu}(\mathbf{x}_t,\mathbf{x_0},t)$ and $\sigma^2(t)$ are mean and covariance of $q(\mathbf{x_{t-1}}|\mathbf{x}_t,\mathbf{x_0})$. $\sigma^2(t)$ can be expressed as:
\begin{equation}
 \sigma^2(t)=\tfrac{(1-\alpha_t)(1-\Bar{\alpha}_{t-1})}{1-\Bar{\alpha}_{t}},
\label{sigma}
\end{equation}
where $\alpha_t=1-\beta_t$ and $\Bar{\alpha}_t=\prod_{t=1}^T\alpha_t$. $\hat{\mu}$ can be formulated as follows through parameterization:
\begin{equation}
    \mu_\theta(\mathbf{x}_t,t,\mathbf{y})=\tfrac{1}{\sqrt{\alpha_t}}\left(\mathbf{x}_t-\tfrac{\beta_t}{\sqrt{1-\Bar{\alpha}_t}}\epsilon_\theta(\mathbf{x}_t,t,\mathbf{y})\right),
\end{equation}
where $\epsilon_\theta(\mathbf{x}_t,t,\mathbf{y})$ is a noise approximator which can be inferred by diffusion model, and $\mathbf{y}=\mathbf{a}~\text{or}~\mathbf{b}$ based on user's group. In this work, as shown in Figure \ref{fig1}, we let 
\begin{equation}
    \begin{aligned}
\epsilon_\theta\left(\mathbf{x}_t,t,\mathbf{y}\right)=~&\mathrm{MLP}_3
    \big(\mathrm{Atten}\big(\mathrm{MLP}_2(\mathbf{y}), \mathrm{MLP}_1(\mathbf{x}_t,\mathbf{t}),\\ \mathrm{MLP}_1(\mathbf{x}_t,\mathbf{t})\big)\big)
    =~&\mathrm{MLP}_3
    \Big(\mathcal{C}\big( \mathrm{MLP}_1(\mathbf{x}_t,\mathbf{t}), \mathbf{y}\big)\Big),
    \end{aligned}
\end{equation}
where $\mathbf{t}$ denotes the embedding vector (one-hot encoding) of time step $t$ and $\mathcal{C}$ denotes the counterfactual module. The parameter set $\theta$ contains all parameters of $\mathrm{MLP}_1$, $\mathrm{MLP}_2$, $\mathrm{MLP}_3$, and $\mathrm{Atten}$.
For convenience, we let $\mathbf{G}=[\mathbf{g}_1,\mathbf{g}_2,\ldots,\mathbf{g}_n]$, where $\mathbf{g}_j=\mathbf{a}$ if $j\in B$ and $\mathbf{g}_j=\mathbf{b}$ if $j\in A$. $\mathbf{G}$ is a matrix of counterfactual group vectors. The training process is summarized in Algorithm~\ref{alg1}. Once the training is finished, the unobserved interaction can be predicted by sampling from $\mathbf{x}_T$ iteratively according to the description in Section \ref{sec_pre}. The detailed procedures are summarized in Algorithm~\ref{alg2}. Note that the prediction process has some randomness due to the sampling of $\mathbf{z}$, one may run the algorithm multiple times and use the mean or median as the final prediction for each user, though we find that the standard deviation is often tiny.

\begin{algorithm}
	\renewcommand{\algorithmicrequire}{\textbf{Input:}}
	\renewcommand{\algorithmicensure}{\textbf{Output:}}
	\caption{DifFaiRec Training}
	\label{alg1}
	\begin{algorithmic}[1]
		\REQUIRE $\mathbf{R}$, $\mathbf{M}$, $\mathbf{G}$, $T$, $\eta$.
		\REPEAT
		\STATE Sample a batch of users’ interactions $\mathbf{X} \subset \mathbf{R}$.
            \STATE Encode $\mathbf{a}$ and $\mathbf{b}$ to $\Bar{\mathbf{a}}$ and $\Bar{\mathbf{b}}$.
		\FORALL{$\mathbf{x} \in \mathbf{X}$}
                \STATE Draw the corresponding $\mathbf{m}\in \mathbf{M}$, $\mathbf{y} \in \mathbf{G}$;
                \STATE Sample $t\sim U(1,T)$ and $\epsilon \sim N(\mathbf{0},\mathbf{I})$.
                \STATE Calculate $\mathbf{x}_t$ and $\mathbf{t}$.
                \STATE Attention key and value: $\mathbf{z}=\mathrm{MLP}_1(\mathbf{x}_t,\mathbf{t})$.
                \STATE Attention query: $\bar{\mathbf{g}}=\mathrm{MLP}_2(\mathbf{y})$.
                \STATE Calculate $\epsilon_\theta\left(\mathbf{x}_t,t,\mathbf{y}\right)=\mathrm{MLP}_3(\mathrm{Atten}(\bar{\mathbf{g}},\mathbf{z},\mathbf{z}))$.
                \STATE Gradient descent: $\theta\leftarrow\theta-\eta\nabla_\theta \left\Vert \mathbf{m} \odot (\epsilon- \epsilon_\theta) \right\Vert^2$.
            \ENDFOR
		\UNTIL Convergence criterion is met.
		\ENSURE Optimized $\theta$.
	\end{algorithmic}  
\end{algorithm}

\begin{algorithm}
	\renewcommand{\algorithmicrequire}{\textbf{Input:}}
	\renewcommand{\algorithmicensure}{\textbf{Output:}}
	\caption{Rating Prediction (for each user)}
	\label{alg2}
	\begin{algorithmic}[1]
		\REQUIRE $\epsilon_\theta$ (given by Algorithm \ref{alg1}), $\{\beta_{t}\}_{t=1}^T$, $x_T$, $T$.
		\FOR{$t=T,\ldots,2,1$}
                \STATE Calculate $\alpha_t$, $\Bar{\alpha}_t$ and $\sigma^2$ based on \eqref{sigma}.
                \STATE Sample $z \sim N(\mathbf{0},\mathbf{I})$.
                \STATE Calculate $\mathbf{x}_{t-1}$ based on \eqref{sampling}.
            \ENDFOR
		\ENSURE Reconstructed $\mathbf{x}_0$.
	\end{algorithmic}  
\end{algorithm}

\subsection{Essential Mathematical Analysis}
Here, we stand on Bayesian thinking to explain why our model works. For the convenience of analysis, we first introduce several events. Denote the counterfactual world as $Z$, where the users are counterfactual users. $I$ denotes rating event for an arbitrary item, $\epsilon$ denotes sampling noise event in forward process, and $\epsilon_\theta$ denotes inferring noise event. 

If the recommender is fair (this paper focuses on the group fairness mentioned in Section 4.1.), sensitive features should not affect the score predicted by the model. That is, the user will still give the same rating in the counterfactual world that can be abstracted into the following formula:
\begin{equation}
    P(I|Z)=P(I),
\end{equation}
where $P(I|Z)=\frac{P(Z,I)}{P(Z)}$ is a conditional probability.

Consequently, the objective for DifFaiRec is to: 1) minimize the distance between $\epsilon$ and $\epsilon_\theta$; 2) make the two events $Z$ and $I$ as independent as possible which can be expressed as:
\begin{equation}
\left\{ 
\begin{aligned}
    &P(\epsilon)=P(\epsilon_\theta),  \\
    &P(Z,I)=P(Z)P(I). 
    \end{aligned}  
\right.
\label{eq17}
\end{equation}

The rating distribution in the real world is $P(I|Z=s)$ and the rating distribution in the counterfactual world is $P(I|Z=1-s)$, where $s\in{0,1}$ denotes the sensitive attribute and $Z=1-s$ is conducted by the counterfactual module discussed in Section 4.3. If \eqref{eq17} holds, the model is fair, i.e., $P(Z,I)=P(Z)P(I)$, meaning
\begin{equation}
    P(I|Z=s)=P(I|Z=1-s).
    \label{eq-fair}
\end{equation}

Estimating $\epsilon$ is equivalent to estimating $I$ from noise. In our diffusion model, $\epsilon$ is performed on the data related to $Z=s$ while the input for $\epsilon_\theta$ 
 is related to $Z=1-s$. Thus, minimizing the difference between $P(\epsilon)$ and $P(\epsilon_\theta)$ is equivalent to minimizing the difference between $P(I|Z=s)$ and $P(I|Z=1-s)$. This means that the second optimization objective is part of the first one, and the second objective constrains the first one implicitly. Further, the two objectives in the entire training can be reduced to:
\begin{equation}
P\left(\prod_{t=1}^T\epsilon_\theta\right)=P\left(\prod_{t=1}^T\epsilon\right),
    \label{eq18}
\end{equation}
which can be optimized with forward and reverse processes by diffusion model iteratively. 
In other words, due to the introduction of the counterfactual module, the model only needs to focus on the rating reconstruction task, which is naturally constrained by fairness items.

Because the diffusion model undergoes the aforementioned Markov process when reconstructing user feedback, we can assume that the two objectives in \eqref{eq17} are reduced to one objective in \eqref{eq18}. Therefore, the above derivation holds only for diffusion models while not for other generative models (\textit{e.g.} GANs and VAEs). Therefore, we prove that the diffusion model has the advantage of generating fair recommendation results compared to other generative recommenders.

Which of PCA and mean pooling can better represent the counterfactual world? Mean pooling retains the mainstream information of the world, and the information contained would not have a too large variance. However, PCA needs to retain as much information as possible, and the data is mapped to a plane with greater variance. Due to the retention of more information, the model with PCA might get a more accurate recommendation performance, because more information can help the model mine user interests more efficiently. In addition, PCA has a feature crossover ability compared to mean pooling. However, especially in the recommendation scenario, the data will have a relatively serious long-tail distribution, which means that retaining more user information may introduce some noise, resulting in a shift in the representation of the main information of the counterfactual world, affecting the fairness of the model. 

\section{Experiments}
\subsection{Experimental Settings}
\subsubsection{Datasets} 
We consider two public datasets, MovieLens-1M\footnote{https://grouplens.org/datasets/movielens/} and LastFM\footnote{http://www.lastfm.com}, which are widely used to evaluate recommendation algorithms. 

     \textbf{MovieLens-1M} contains 1 million ratings on 3,900 movies given by 6,040 users. The ratings are made on a 5-star scale and each user has at least 20 ratings. Since this dataset contains the gender and age attributes of the users, we chose these two attributes as the protected attributes. In terms of age, the users are divided into young people and old people by 50.
    
     \textbf{LastFM} is a dataset of music recommendations. For each user, there is a list of his or her favorite artists along with the number of plays. Here, we take the number of plays as the rating. Since this dataset lacks user information, we divide the users into two groups according to their activity level and interest diversity. Here, a user-artist interaction is defined as the user has listened to the artist’s songs. The play is defined as the number of times a user plays an artist’s songs and the given tag represents various music styles. The users are grouped as active or inactive users by an activity level boundary of 15,000 plays and are grouped as users with focused or divergent interests by an interest diversity boundary of 300 tags associated with the artists.  
The statistics of the two datasets and their groups are shown in Table~\ref{tab1}.

\begin{table}[htbp]
  \centering
  \caption{Statistics of the two datasets.}
  \resizebox{\linewidth}{!}{
    \begin{tabular}{c|cccc}
    \hline
          & user \& protected attribute  & item  & interaction & \multicolumn{1}{l}{sparsity} \\
    \hline
    MovieLens & 6,040 \makecell{(male 71.7\%, female 28.3\%)\\(young 85.5\%, old 14.5\%)} & 3,900 & 1,000,209 & 0.0425 \\ \hline
    LastFM & 1,892 \makecell{(active 63\%, inactive 37\%)\\(focused 9.5\%, divergent 90.5\%)} & 17,632 & 92,834 & 0.0028 \\
    \hline
    \end{tabular}}%
  \label{tab1}%
\end{table}%


\subsubsection{Baselines}
We compare DifFaiRec with six baselines including two utility-focused baselines and four fairness-focused baselines. The details are as follows.

\textbf{MF} \cite{rendle2012bpr} is a well-known collaborative filtering technique based on matrix factorization. 
    \textbf{AutoRec} \cite{sedhain2015autorec} is one of the earliest neural network or deep learning based recommendation models. 
   \textbf{IFGAN}  \cite{lin2021info} is a GAN4Rec model. Different from traditional GANs, this model employs two generators and uses the gradient information between the generator and discriminator to leverage the sampling strategy combining the two generators. One generator is responsible for generating hard negative samples, the other generator is responsible for generating hard sample pairs, and the discriminator is responsible for distinguishing the sample pairs.
     \textbf{DiffRec} \cite{wang2023diffusion} is the first diffusion-based recommender that takes the diffusion model as the backbone to generate the user's feedback.  \textbf{ChatGPT4Rec} \cite{liu2023chatgpt} is based on large language models that can transform the recommendation task into a natural language generation by constructing a specific prompt.
   \textbf{FairC} \cite{bose2019compositional} is an adversarial framework to train filters to get graph embeddings without information about the users' protected attributes. It can enforce fairness based on different combinations of fairness constraints.
   \textbf{FairGO} \cite{wu2021learning} presents a combination structure of sub-filters aiming to remove the information of sensitive attributes. The model is trained via an adversarial technique on a user-centric graph.
    \textbf{FairGAN} \cite{li2022fairgan} was originally proposed for exposure fairness, which, however, can be regarded as group fairness on the item or user side. Thus we compare FairGAN in our experiments. 

\subsubsection{Evaluation Metrics}
We use three popular metrics including Recall (R@k), Normalized Discounted Cumulative Gain (N@k), Absolute Equality (A@k), and Equal Opportunity (E@k), to evaluate all methods considered in this paper. Particularly, A@k and E@k are defined as
\begin{equation}
    \text{A}@k:=|u_A-u_B|,
\end{equation}
where $u_A$ and $u_B$ are mean absolute errors (MAEs) for groups A and B.
\begin{equation}
\text{E}@k:=\sqrt{|e_A-e_B|},
\end{equation}
where $e_A$ is the ratio of the number of items that appear in the top-k list incorrectly to the total number of items in the list in group level for group A and $e_B$ is that for group B.

The larger R@k and N@k indicate better recommendation performance while the smaller A@k and E@k mean the fairer recommendations.

\subsubsection{Parameter Settings}
In our DifFaiRec, we set $T$ (the number of diffusion steps or inference steps) as 100. Note that increasing $T$ may improve the performance but the gain is not significant.  The size of step embedding is 64. The lower bound and upper bound of the variance $\beta$ is set to $\text{sigmoid}(-6)\times5e^{-4}+1e^{-5}$ and $\text{sigmoid}(6)\times5e^{-4}+1e^{-5}$ respectively, where the variance scale $L$ is $1e^{-4}$. In the training process, the batch size is set to $64$ and the initial learning rate is $1e^{-3}$ with an adaptive moment estimation (Adam) \cite{kingma2014adam} optimizer. Besides, the $k$ for top-$k$ metrics are set to $7$ and $10$ in the MovieLens dataset and LastFM dataset, respectively. The datasets are split with 8:1:1 to validate the performance of methods and we ensure that each user has at least 10 observed interactions in the training set to guarantee the reasonableness of evaluation. All experiments are implemented using a single Tesla-V100 GPU. 

\subsection{Comparison with Baselines}
\begin{table*}[t]
  \centering
  \caption{Comparison with baselines on MovieLens. The best results are presented in bold font. All the results are highly statistically significant ($p$-value<0.01) under paired t-tests.}
    \begin{tabular}{cccccccccc}
    \hline
    \multicolumn{10}{c}{MovieLens} \\
    \hline
          &       & \multicolumn{4}{c}{age} & \multicolumn{4}{c}{gender} \\
    \hline
          & models & recall↑ & ndcg↑ & Abs equality↓ & equal O↓ & recall↑ & ndcg↑ & Abs equality↓ & equal O↓ \\
    \hline
    {typical baselines} & MF    & 0.233  & 0.397 & 0.074 & 0.301  & 0.236  & 0.401 & 0.079  & 0.269  \\
          & AutoRec & 0.138  & 0.361 & 0.073  & 0.332  & 0.141  & 0.356 & 0.076  & 0.274  \\
          & IFGAN & 0.394  & 0.543 & 0.052 & 0.298  & 0.401  & 0.551 & 0.056  & 0.237  \\
          & DiffRec & 0.397 & 0.540 & 0.049 & 0.307 & 0.436 & 0.549 & 0.052 & 0.241\\
          & ChatGPT4Rec & 0.286 & 0.399 & 0.047 & 0.301 & 0.292 & 0.395 & 0.054 & 0.239\\
    \hline
    {fair baselines} & FairC & 0.391  & 0.537 & 0.034  & 0.286  & 0.434  & 0.543 &0.031  & 0.226  \\
          & FairGO & 0.379  & 0.444 & 0.030 & 0.277  & 0.398  & 0.476 & 0.027 & 0.221  \\
          & FairGAN & 0.397  & 0.541 & 0.019 & 0.267  & 0.438  & 0.547& 0.022 & 0.219  \\
    \hline
    {proposed methods} & DifFaiRec1 & 0.400  & 0.549 & \textbf{0.016} & \textbf{0.261 } & 0.442  & 0.556 & 0.020  & \textbf{0.216 } \\
          & DifFaiRec2 & \textbf{0.401 } & \textbf{0.551 } & \textbf{0.016} & 0.263  & \textbf{0.446 } & \textbf{0.559 } & \textbf{0.019} & 0.221  \\
    improvement &       & 1.01\% & 1.85\% & 15.79\% & 2.25\% & 1.83\% & 1.45\% & 13.64\% & 1.37\% \\
    \hline
    \end{tabular}%
  \label{tab2}%
\end{table*}%

\begin{table*}[t]
  \centering
  \caption{Comparison with baselines on LastFM. The best results are presented in bold font. All the results are highly statistically significant ($p$-value<0.01) under paired t-tests.}
    \begin{tabular}{cccccccccc}
    \hline
    \multicolumn{10}{c}{LastFM} \\
    \hline
          &       & \multicolumn{4}{c}{activity level} & \multicolumn{4}{c}{interest diversity} \\
    \hline
          & {models} & recall↑ & ndcg↑ & Abs equality↓ & equal O↓ & recall↑ & ndcg↑ & Abs equality↓ & equal O↓ \\
    \hline
    \multicolumn{1}{c}{{typical baselines}} & \multicolumn{1}{c}{MF} & 0.063  & 0.147 & 0.717 & 0.112  & 0.061  & 0.141 & 0.723 & 0.127  \\
          & \multicolumn{1}{c}{AutoRec} & 0.102  & 0.152 & 0.691 & 0.108  & 0.099  & 0.153 & 0.700 & 0.112  \\
          & \multicolumn{1}{c}{IFGAN} & 0.114  & 0.166 & 0.663 & 0.101  & 0.112  & 0.159 & 0.668 & 0.110  \\
          & DiffRec & 0.122 & 0.166 & 0.679 & 0.100 & 0.129 & 0.171 & 0.693 & 0.112\\
          & ChatGPT4Rec & 0.119 & 0.142 & 0.647 & 0.101 & 0.116 & 0.139 & 0.662 & 0.108\\
    \hline
    \multicolumn{1}{c}{{fair baselines}} & \multicolumn{1}{c}{FairC} & 0.118  & 0.144 & 0.581 & 0.094  & 0.116  & 0.143 & 0.587 & 0.103  \\
          & \multicolumn{1}{c}{FairGO} & 0.120  & 0.154 & 0.576 & 0.089  & 0.125  & 0.160 & 0.579 & 0.104  \\
          & \multicolumn{1}{c}{FairGAN} & 0.124  & 0.168 & 0.577 & 0.081  & 0.131  & 0.175 & 0.581 & 0.099  \\
    \hline
    \multicolumn{1}{c}{{proposed methods}} & \multicolumn{1}{c}{DifFaiRec1} & 0.124  & 0.171 & 0.557 & \textbf{0.080 } & 0.133  & 0.180 & 0.564 & \textbf{0.097 } \\
          & \multicolumn{1}{c}{DifFaiRec2} & \textbf{0.128 } & \textbf{0.174 } & \textbf{0.555} & 0.082  & \textbf{0.135 } & \textbf{0.184 } & \textbf{0.563} & 0.102  \\
    improvement &       & 3.23\% & 3.57\% & 3.65\% & 1.23\% & 3.05\% & 5.14\% & 2.76\% & 2.02\%\\
    \hline
    \end{tabular}%
  \label{tab3}%
\end{table*}%

\textbf{Effectiveness of DifFaiRec.} The comparison results are shown in Table~\ref{tab2} and \ref{tab3}. The best results are bold-faced. DifFaiRec1 is a model with the mean pooling group vector builder while DifFaiRec2 is a model with the PCA group vector builder. The results indicate that DifFaiRec outperforms both utility-focused baselines and fairness-aware baselines on recommendation performance on all datasets. Specifically, our proposed method exhibits improvement 1.01\%, 1.85\%, 15.79\%, 2.25\% on age dimension and 1.83\%, 1.45\%, 13.64\%, 1.37\% on gender dimension in terms of recall, ndcg, absolute equality, and equal opportunity on MovieLens dataset and 3.23\%, 3.57\%, 3.65\%, 1.23\% on activity level dimension and 3.05\%, 5.14\%, 2.76\%, 2.02\% on interest diversity dimension in terms of recall, ndcg, absolute equality, and equal opportunity on LastFM dataset, respectively. This illustrates the effectiveness of the proposed fairness-aware recommendation algorithm on extracting user preference under fairness-aware constraints with the help of the powerful fitting ability of the diffusion model and the decoupling strategy for rating prediction and fairness keeping of the counterfactual module. Interestingly, we notice that although adding a fairness constraint may cause some items to be poorly recommended, it may have an overall positive benefit for recommendation quality in the big picture. Therefore, the recommendation performance could be further improved with fairness concerns. More significantly, the better fairness of DifFaiRec validates the capability of the counterfactual module in the fairness representation task and the flexibility of the conditional diffusion model.

\textbf{Difference between DifFaiRec1 and 2.} The only difference between these two models is the group vector builder. As in the previous analysis, PCA can retain more information of sample distribution and have a certain degree of feature interaction themselves. However, due to the lack of redundant information or long-tail information, the counterfactual constraints may be weakened, resulting in the final fairness of the model being inferior to that of the mean pooling version.

\subsection{Ablation Study}
\begin{figure}[htbp]
	\centering
	\subfloat[Ablation study in terms of ndcg.]{\includegraphics[width=.48\columnwidth]{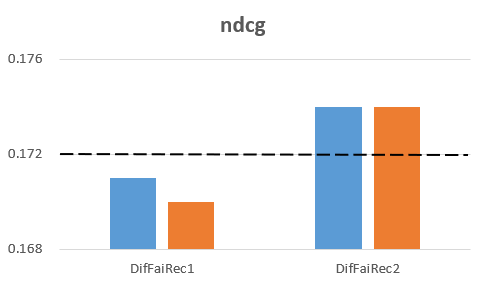}}\hspace{2pt}
	\subfloat[Ablation study in terms of equal opportunity.]{\includegraphics[width=.48\columnwidth]{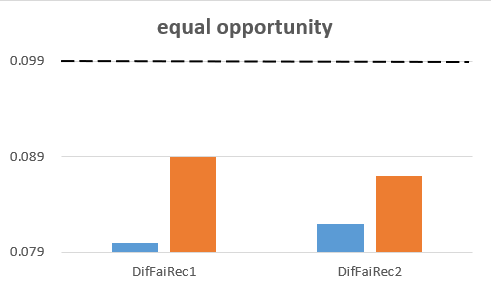}}\\
 \caption{Ablation study. The blue bar draws the performance of the proposed model and the orange bar draws the performance of the model without using the condition encoder $\mathrm{Enc}$ (i.e. $\mathrm{MLP}_2$). Additionally, the black dotted line illustrates the performance of the model without the counterfactual module.}
 \label{fig2}
\end{figure}

To explore the effectiveness of our proposed counterfactual module for keeping fairness, we conduct ablation experiments on it. We explore the role of the condition encoder and counterfactual module. The experiment is conducted on the LastFM dataset grouped with activity level. In Fig.~\ref{fig2}, the blue bar draws the performance of the proposed model and the orange bar draws the performance of the model with the condition encoder detached. Additionally, the black dotted line illustrates the performance of the model without the counterfactual module. As mentioned above, the condition encoder improves the counterfactual representation quality, which in turn enforces the fairness constraints of the model. After removing the counterfactual module, the fairness of the model is heavily reduced while the fluctuation of ndcg is small. It shows that our proposed counterfactual module can effectively improve the fairness of the model and ensure the recommendation quality.

\begin{figure}[t]
	\centering
	\subfloat[Effects of diffusion step.]{\includegraphics[width=.48\columnwidth]{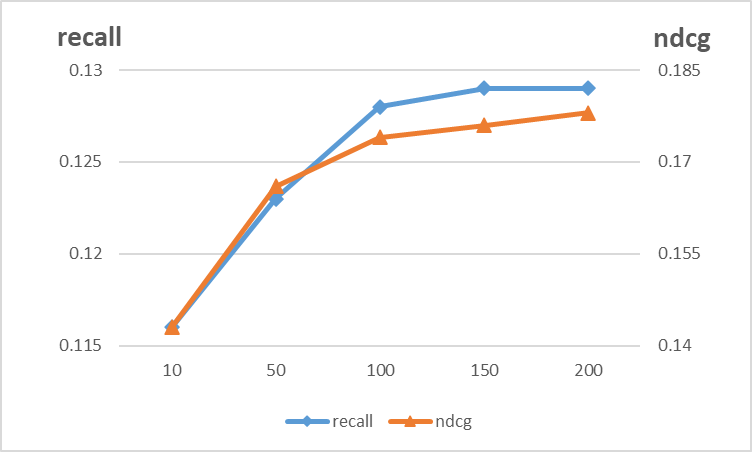}}\hspace{2pt}
	\subfloat[Effects of variance scale.]{\includegraphics[width=.48\columnwidth]{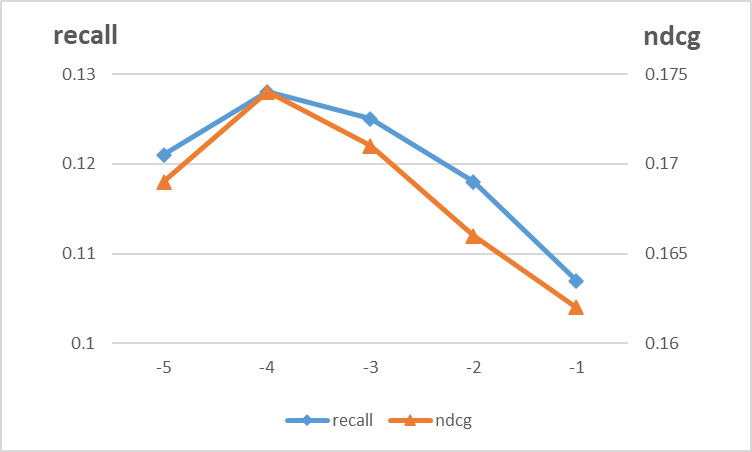}}\\
 \caption{Results of impact of hyper-parameters. The blue line represents the variation of recall and the orange line represents the variation of ndcg.}
 \label{fig3}
\end{figure}
\subsection{Impact of Hyper-parameters}
In DifFaiRec, there are two important hyper-parameters: diffusion step $T$ and variance scale $L$. To illustrate their impacts, we vary $T$ from 10 to 200 and change $L$ from 1e-5 to 1e-1, respectively. We take DifFaiRec with PCA (DifFaiRec2) to do experiments on the LastFM dataset grouped with activity level. It can be seen from Fig.~\ref{fig3} that the recommendation performance increases as the diffusion step becomes larger. But after 100 steps, the ascension is very limited. At the same time, the model with a large diffusion step is time-consuming. Thus, we set the diffusion step to 100. When it comes to the variance scale, the performance first increases followed by a sharp reduction which might be that the excessive noise destroys the basic assumptions of the diffusion model, resulting in deviations in the sample reconstruction process, making the recommendation poor. Therefore, using  DifFaiRec requires careful choice of variance scale.

\begin{table}[htbp]
  \centering
  
  \caption{The effect of group sparsity on DifFaiRec2.}
  \resizebox{\linewidth}{!}{
    \begin{tabular}{c|cccc|cccc}
    \hline
     & \multicolumn{4}{c|}{M-G} & \multicolumn{4}{c}{L-ID} \\
    \hline
     SR & recall & ndcg & A. equality & equal O & recall & ndcg & A. equality & equal O \\
    \hline
    50\% & 0.441 & 0.553& 0.022& 0.224& 0.130& 0.176& 0.578& 0.106\\
    70\% &0.437& 0.549 & 0.029 & 0.226 & 0.128 & 0.175 & 0.670& 0.110\\
    90\%& 0.430& 0.544& 0.051& 0.238& 0.119& 0.168& 0.691& 0.112\\
    \hline
    \end{tabular}}%
  \label{tab-sparse}%
\end{table}%

\subsection{Robustness Study for Group Sparsity}
The performance of our model depends on the quality of the group vector, which is also the limitation of our approach. According to the law of large numbers, the sparsity within a group (the number of users in a group) affects the quality of the group vector. Therefore, we conducted experiments on the effect of group sparsity on the model.

Table~\ref{tab-sparse} shows the results of the group sparsity test for DifFaiRec2 on MovieLens gender level (M-G) and LastFM interest diversity level (L-ID). We randomly under-sample the users in the minority group (female in M-G and focused interests in L-ID), and the sampling ratio (SR) is set to {50\%, 70\%, 90\%}. As the number of users in the minority group decreases (the minority group becomes sparser), the accuracy metrics change little, but the fairness metrics change significantly. In the extreme case (90\% under-sampled), the fairness metrics become the level of the unfairness baseline. This shows that our fairness method no longer works well under extremely sparse conditions. Fortunately, this situation is very rare, and most of the time our method works and is robust to sparse.

\section{Conclusion}
This paper proposes a diffusion model based fair recommender, called DifFaiRec addressing the fairness issues in recommendations via an attention-based counterfactual module. To keep fairness, two group vector building methodologies are proposed to find the difference between the two groups. The proposed DifFaiRec generates a fair recommendation list considering group fairness on different sensitive features while maintaining user utility as well as possible. In particular, we give a mathematical analysis for our proposed method. Finally, extensive experiments show the advantages of the proposed recommender over the state-of-the-art baselines. In our future work, we are interested in investigating the issues of fairness across both user and item sides, which are vital for recommendations as well.

\section*{Acknowledgements}

This work was supported by the National Natural Science Foundation of China under Grants No.62106211 and No.62376236,  the General Program of Natural Science Foundation of Guangdong Province under Grant No.2024A1515011771, the Guangdong Provincial Key Laboratory of Mathematical Foundations for Artificial Intelligence (2023B1212010001), Shenzhen Science and Technology Program ZDSYS20230626091302006, and Shenzhen Stability Science Program 2023.

\bibliographystyle{IEEEtranS.bst}
\bibliography{ref.bib}

\appendix

\subsection{More explanation for Section 4.5}

$\epsilon$ is independent for $I$ but $\epsilon_\theta$ depends on $I$. There is no conflict. To clarify this, let's first consider the following facts.
\begin{itemize}
    \item $\epsilon$ is the noise sampled randomly from a Gaussian distribution and hence is independent of $I$.
    \item $\epsilon_\theta$ is the output given by the diffusion model (DM), and the input of the DM includes $I$. Thus, $\epsilon_\theta$ depends on I.
    \item $P(\epsilon)=P(\epsilon_\theta)$ is an ideal state. In reality, we are only able to approximate $P(\epsilon)$ using $P(\epsilon_\theta)$, that means $P(\epsilon)\approx P(\epsilon_\theta)$.
\end{itemize}

In conclusion, there is no conflict.

In the diffusion model, recovering the noise $\epsilon$ is equivalent to recovering the data $x_0$. If $\epsilon \approx \epsilon_\theta$, then $I \approx I_\theta$. We know that $I$
is based on $Z=s$ and $I_\theta$ is based on $Z=1-s$. Then $I$ and $I_\theta$ should be independent from $Z$ approximately. Thus, $I_\theta$ is fair approximately.

\subsection{Experimental Settings of Baselines}

For all baselines, the batch size is set to 64 and the learning rate is set to $1e^{-3}$. In MF, the weight of regularisation term is set to 0.1 and the hidden size of factorized vector is set to 20. In AutoRec, the weight of regularisation term is set to 0.1 and the hidden size is set to 500. In IFGAN, the number of units in hidden layers in generator is [1024, 256, 1024] and in discriminator is [1024, 256]. In DiffRec, the step embedding size is fixed at 10, the noise scale is set to $1e^{-5}$ and the upper bound of noise is set to 
$5e^{-5}$. The number of units in hidden layers is [600, 200]. The diffusion step is set to 100 and the reverse step is set to 50. In FairC, the number of units in hidden layers in discriminators and filters are [1024, 512, 256] and the parameter for fairness is set to 0.1. In FairGO, the filtered embedding size is set as 64. The number of units in hidden layers is [128, 64]. The balance parameter for fairness is set to 0.1. In FairGAN, the number of units in hidden layers is [500, 500, 500]. The penalty coefficient is set to 0.1 and the weight parameter of fairness is set to 0.1.

\end{document}